# Large-Area Photonic Membranes Achieving Uniform and Strong Enhancement of Photoluminescence and Second-Harmonic Generation in Monolayer WSe$_2$

*‡Fong-Liang Hsieh, ‡Chih-Zong Deng, Shao-Ku Huang, Tsung-Hsin Liu, Chun-Hao Chiang, Che-Lun Lee, Man-Hong Lai, Jui-Han Fu, Vincent Tung, Yu-Ming Chang, Chun-Wei Chen, and Ya-Lun Ho*

F. L. Hsieh, C. Z. Deng, C. H. Chiang, Y. L. Ho
Research Center for Electronic and Optical Materials, National Institute for Materials Science (NIMS), 1-1 Namiki, Tsukuba, Ibaraki 305-0044, Japan
E-mail: Ya-Lun Ho, HO.Ya-Lun@nims.go.jp

F. L. Hsieh, S. K. Huang, T. H. Liu, C. H. Chiang, Che-Lun Lee, C. W. Chen
Department of Materials Science and Engineering, National Taiwan University, No. 1, Sec. 4, Roosevelt Rd., Taipei 10617, Taiwan
E-mail: Chun-Wei Chen, chunwei@ntu.edu.tw

Man-Hong Lai, Y. M. Chang
Center for Condensed Matter Sciences, National Taiwan University, No. 1, Sec. 4, Roosevelt Rd., Taipei 10617, Taiwan

J. H. Fu, V. Tung
Department of Chemical System Engineering, School of Engineering, The University of Tokyo, 7-3-1 Hongo, Bunkyo, Tokyo 113-8656, Japan

Y. M. Chang, C. W. Chen
Center of Atomic Initiative for New Materials (AI-MAT), National Taiwan University, No. 1, Sec. 4, Roosevelt Rd., Taipei 10617, Taiwan

Funding: JSPS KAKENHI Grant Number JP23K26155, JP25KF0083, Advanced Research Infrastructure for Materials and Nanotechnology in Japan (ARIM) of the Ministry of Education, Culture, Sports, Science and Technology (MEXT) (Proposal Number 25NM5090), Ministry of Education in Taiwan (Grant Number 111L900801)

**Two-dimensional (2D) transition metal dichalcogenides (TMDs) exhibit strong excitonic responses, direct bandgaps, and remarkable nonlinear optical properties, making them highly attractive for integrated photonic, optoelectronic, and quantum applications. Here, we present a large-area freestanding membrane photonic platform that achieves exceptional enhancement of light–matter interactions in monolayer WSe$_2$ via quasi-bound states in the continuum (quasi-BICs). The freestanding architecture effectively suppresses radiative losses and supports high-Q optical resonances, leading to enhanced light–matter interactions. This results in significant photoluminescence (PL) emission and second-harmonic generation (SHG) enhancement factors of 1158 and 378, respectively, with spatial uniformity sustained across a 450 × 450 μm$^2$ area. This uniform SHG enhancement further enables polarization-resolved mapping of crystal orientation and grain boundaries, offering a practical method for large-area structural characterization of 2D materials. Moreover, femtosecond-pumped SHG spectra reveal multiple narrowband peaks originating from distinct quasi-BIC modes—providing direct spectral evidence of resonantly enhanced nonlinear coupling. The combined attributes of strong optical enhancement, spectral selectivity, and wafer-scale compatibility establish this platform as a scalable interface for 2D semiconductor integration in next-generation optoelectronic, nonlinear, and quantum photonic technologies.**

‡F. L. Hsieh and C. Z. Deng contributed equally to this work.

# 1. Introduction

Two-dimensional (2D) semiconductor transition metal dichalcogenides (TMDs) have emerged as a promising material system for optoelectronic applications due to their distinctive optical and electronic properties. These 2D materials exhibit direct bandgaps spanning the visible to near-infrared spectrum, and possess strong light–matter coupling, making them highly attractive for future quantum devices and integrated photonic systems.[1-5] Despite these advantages, monolayer TMDs often suffer from low optical responses, including low emission and absorption efficiency, resulting from their atomically thin nature. This limitation restricts their utility in photonic and quantum applications. To overcome these challenges and fully exploit their potential in practical devices, substantial efforts have focused on enhancing and engineering their optical response through integration with nanophotonic structures.

Nanophotonic structures, including independent dielectric/metallic nanocavities and photonic-crystal cavities, as well as extended-area periodic photonic structures such as metasurfaces, offer powerful means to enhance light confinement and to control the emission intensity and directionality of light emitted from TMDs.[6-40] These structures realize their functionalities through distinct optical resonances. Independent dielectric nanocavities support Mie resonances, while metallic nanocavities exhibit localized surface plasmon resonances. These resonances benefit from ultra-small mode volumes and high Purcell factors, enabling significant enhancement of light–matter interaction characteristics in monolayer TMDs, including photoluminescence (PL), nonlinear optical effects, exciton-polariton emission, and photocurrent generation.[23-34] However, the highly localized nature of these resonant hotspots limits scalability and hinders integration into practical large-area devices. To address these limitations, periodic resonant photonic structures have emerged as promising alternative photonic platforms.[6-22] By supporting collective optical resonances over extended areas, these platforms enable spatially uniform and wavelength-tunable enhancement of these optical responses across continuous large-area TMD monolayers, offering a pathway toward scalable integration.

To advance the enhancement of light–matter interactions between 2D materials and photonic platforms, it is advantageous to incorporate resonant modes with exceptional light confinement. In this context, metasurfaces designed to support bound states in the continuum (BICs) are particularly appealing. BICs arise from the interference of multiple radiative pathways, yielding optical modes that are symmetry-protected or interference-induced, with theoretically infinite Q-factors and negligible radiative losses.[41, 42] These properties enable extreme electromagnetic field confinement, which is crucial for maximizing the interaction between light and 2D TMDs.[9, 15, 18-20, 35-37] Thus, large-area photonic platforms engineered to support BIC resonances offer a compelling strategy to achieve both strong optical enhancement and practical scalability, effectively overcoming the limitations of independent nanocavity-based aapproaches.

BIC-based nanophotonic platforms have predominantly been implemented using substrate-supported structures consisting of high-index materials on low-index substrates. However, this geometry inherently breaks vertical symmetry and introduces radiative leakage into the substrate, thereby weakening field confinement. Freestanding membrane structures overcome these limitations by entirely removing the substrate, restoring vertical symmetry and thereby eliminating substrate-induced radiation losses.[43-45] This suspended configuration not only preserves the low-radiative-loss nature of BICs but also supports strong in-plane and out-of-plane field localization, significantly enhancing light–matter interactions. Moreover, the absence of substrate-related perturbations leads to a broader range of tunability in the BIC resonances, facilitating precise spectral engineering. The enhanced mode confinement in the freestanding geometry also improves spatial mode overlapping with atomically thin layers. This promotes more efficient coupling to 2D materials, making them ideal architecture for realizing 2D-material-coupled photonic platforms.[18, 21, 46]

Here, we present a freestanding and large-area membrane photonic platform specifically designed for monolayer TMDs, through quasi-BIC-enhanced resonances enabling exceptional light–matter interactions across large areas. Owing to its suspended membrane architecture, our platform achieves a PL enhancement factor of 1158, which is among the highest reported for periodic photonic structures. This continuous and spatially homogeneous enhancement is maintained over a 450 × 450 μm² area, representing a large active region for PL amplification in monolayer $WSe_2$. Beyond enhancing PL emission efficiency, SHG measurements reveal multiple clearly resolved and narrow-linewidth peaks arising from distinct quasi-BIC resonances. This result demonstrates unprecedented multiband nonlinear spectral clarity not previously reported. Moreover, this large-area SHG enhancement enables polarization-resolved SHG mapping of crystal orientation and grain boundaries in monolayer $WSe_2$ over extended areas. This capability provides a powerful approach for spatially extensive characterization of 2D materials. This membrane photonic platform uniquely combines robust field enhancement, high quality factors, broad spectral tunability, and large-area scalability and uniformity. These combined features establish it as an effective and practical nanophotonic interface for advancing light–matter interactions in 2D semiconductors, paving the way for their integration into wafer-scale photonic and quantum devices.

**2. Large-Area Growth and Scalable Transfer of Monolayer WSe$_2$**

Figure 1 illustrates the complete sample preparation of a membrane photonic platform specifically designed for enhancing light-matter coupling in monolayer WSe$_2$ and presents initial characterizations of the as-grown monolayer WSe$_2$. **Figure 1**a shows a schematic overview, and the design consists of a freestanding silicon nitride (SiN) hole-array membrane as the photonic platform with the monolayer WSe$_2$ transferred on top. This platform is engineered to support BICs, providing strongly enhanced light–matter interactions that are critical for amplifying the optical responses of 2D TMDs. The preparation process is detailed in Figure 1b. A monolayer WSe$_2$ film is first synthesized on a c-plane sapphire substrate using alkali-assisted chemical vapor deposition (AACVD), ensuring large-area growth with high crystallinity.[47] The monolayer WSe$_2$ is then transferred onto a prefabricated SiN membrane featuring a periodic hole-array structure. This is achieved through a polymer-assisted wet transfer technique using ethylene-vinyl acetate (EVA) as the sacrificial carrier layer. This approach enables uniform deposition of a monolayer WSe$_2$ film across the 2-inch wafer, as shown in Figure 1b. Figure 1c presents a sequence of optical microscope (OM) and scanning electron microscope (SEM) images showing key stages of the fabrication process. OM images show the membrane photonic platform at different membrane thicknesses and before and after the transfer of monolayer WSe$_2$. These images clearly reveal the thickness-dependent optical contrast induced by the controlled etching process, as well as the successful and uniform transfer of monolayer WSe$_2$ across the large-area membrane photonic platform surface. The SEM image illustrates the continuous coverage and good morphological conformity of the monolayer WSe$_2$ with the underlying membrane photonic platform. The material quality of the as-grown monolayer WSe$_2$ was characterized in Figure 1d and e. Figure 1d shows an atomic force microscopy (AFM) scan of the film, revealing a smooth surface morphology across the scanned region. The corresponding height profile confirms a monolayer thickness of approximately 0.73 nm and a low surface roughness (Rq ≈ 0.196 nm). The Raman spectrum shown in Figure 1e displays the characteristic peaks of monolayer WSe$_2$, E1 2g (254.1 cm$^{-1}$) corresponding to the in-plane vibration, and A$_{1g}$ (261.1 cm$^{-1}$) representing the out-of-plane vibration of selenium atoms. The PL spectrum as shown in Figure 1f reveals a strong emission peak centered at ~760 nm, confirming the direct bandgap nature and high optical quality of the monolayer WSe$_2$. The two small peaks at 693.8 nm and 695.2 nm are the fluorescence associated with the sapphire with unintentional Cr$^{3+}$ doping.

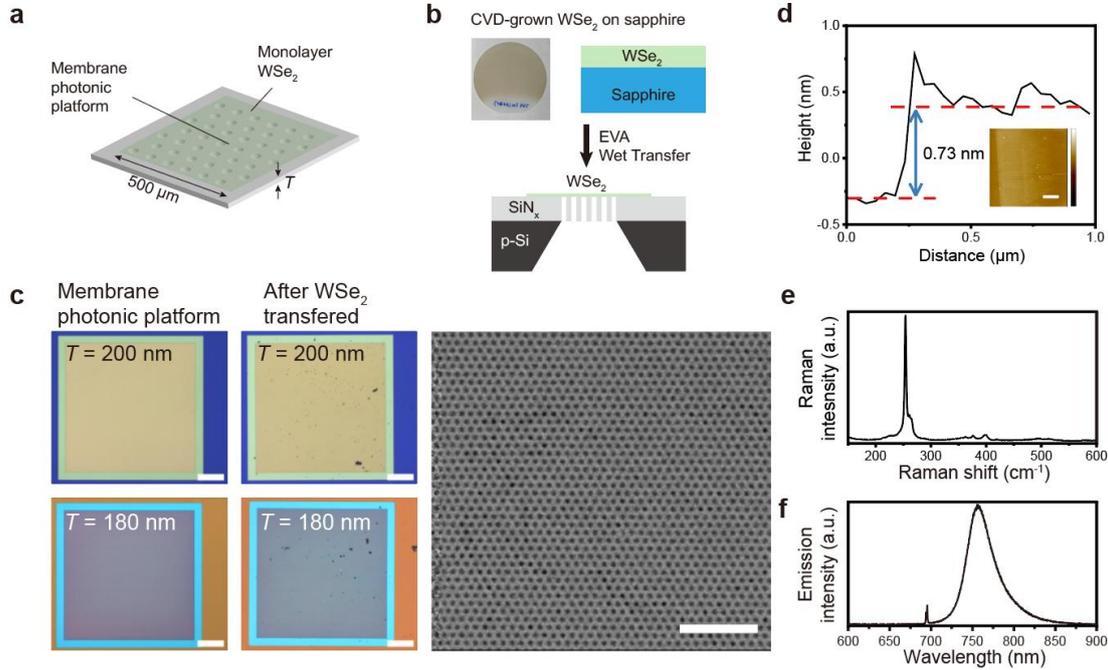

**Figure 1.** Membrane photonic platform tailored to large-area monolayer WSe$_2$ for enhancement of light-matter interaction characteristics. a) Schematic illustration of the large-area membrane photonic platform engineered for monolayer WSe$_2$. b) Fabrication process showing the transfer of CVD-grown monolayer WSe$_2$ onto the membrane photonic platform. The photo image displays the as-grown WSe$_2$ on a 2-inch sapphire wafer. c) OM images of membrane photonic platforms with different membrane thicknesses, before and after monolayer WSe$_2$ transfer, and SEM image of the transferred monolayer WSe$_2$ over the nanostructured membrane. The scale bars are 100 μm for the OM images and 5 μm for the SEM image. d) AFM height profile confirming monolayer thickness of 0.73 nm and surface roughness Rq of 0.196 nm. The inset shows the AFM image of the as-grown monolayer WSe$_2$ with a scale bar of 3 μm and a color scale ranging from -10 to 5 nm. e) Raman spectrum and f) PL spectrum of the as-grown monolayer WSe$_2$ confirming its monolayer nature.

## 3. Design of Quasi-BIC Modes in the Membrane Photonic Platform

The proposed membrane photonic platform consists of a freestanding SiN membrane patterned with a triangular lattice of air holes, onto which a monolayer WSe$_2$ was transferred, as shown in **Figure 2**a. The structure was numerically designed and analyzed using rigorous coupled-wave analysis (RCWA) to simulate its absorption spectra. The structure has a lattice period of 600 nm, a hole diameter of 300 nm, and a variable membrane thickness $T$. The thickness of the WSe$_2$ layer is set to 0.6 nm in the simulation. The incident light is $x$-polarized. Figures 2b and 2c show the simulated absorption as a function of membrane thickness. To focus on symmetry-protected BICs localized at the $\Gamma$ point, the incident angle $\theta$ was kept within 0–5°, as illustrated in Figure S1. Under normal incidence, the optical response features modes with theoretically infinite Q-factors, characteristic of BICs. To excite symmetry-protected BICs, a small incident angle of $\theta = 4°$ was introduced along the $y$- and $x$-directions for Figures 2b and 2c, respectively. The simulated absorption spectra revealing three quasi-BICs (BIC$_1$, BIC$_2$, and BIC$_3$). (Also See Figure S1) The observed thickness-dependent spectral shifts of these modes demonstrate the excellent tunability of proposed membrane photonic platform. A membrane thickness of 215 nm was used for further analysis. At this thickness, the simulated resonance wavelengths of BIC$_1$, BIC$_2$, and BIC$_3$ are 775.0, 784.6, and 800.4 nm, respectively. Figures 2d–f show the electric energy density ($E_{den}$) distributions for BIC$_1$, BIC$_2$, and BIC$_3$. These field profiles show distinct levels of field confinement at the WSe$_2$ monolayer for each mode, with maximum electric energy densities of 486, 1098, and 1820 for BIC$_1$, BIC$_2$, and BIC$_3$, respectively. Figure 2g illustrates the impact of introducing a SiO$_2$ substrate beneath the SiN hole-array slab. The simulations for the membrane and substrate-supported platforms were performed at the same wavelength to ensure identical refractive index and extinction coefficient of WSe$_2$, allowing a direct comparison of field enhancement. The presence of the substrate leads to increased field leakage and significantly reduces the peak field enhancement of BIC$_3$ to approximately 987, nearly half that of the substrate-free membrane configuration, highlighting the advantage of freestanding membrane designs for maximizing light–matter interactions in 2D materials integrated with the membrane photonic platform.

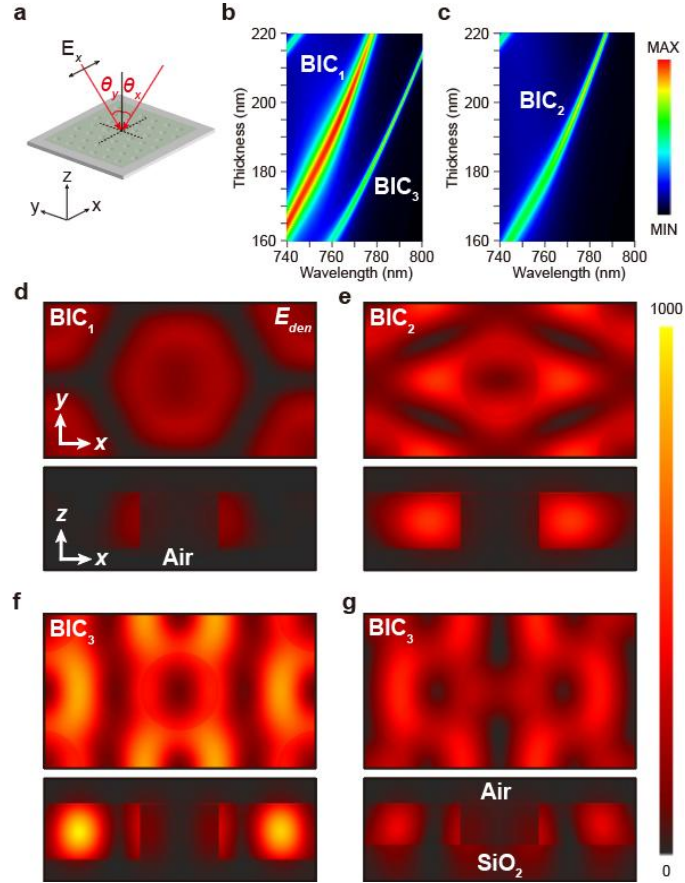

**Figure 2.** Resonance tunability and strong field enhancement in monolayer WSe$_2$ achieved through quasi-BIC resonances in the membrane photonic platform a) Schematic of the simulation modal and b) and c) the simulated absorption spectra as a function of membrane thickness for a 0.6 nm monolayer of WSe$_2$ on a membrane photonic platform, illuminated with x-polarized light at an incident angle of 4°, along the b) y-direction and c) x-direction. d, e, f) Distributions of electric energy density $E_{den}$ at the resonant wavelengths of BIC$_1$, BIC$_2$, and BIC$_3$, respectively, along the xy-plane and the xz-plane. g) Electric energy density $E_{den}$ distribution for BIC$_3$ of the SiN hole-array slab supported by a SiO$_2$ substrate.

## 4. Giant and Uniform PL Enhancement in WSe$_2$ Enabled by the Membrane Photonic Platform

To experimentally evaluate the designed membrane photonic platform, optical characterization was conducted on its coupling with monolayer WSe$_2$. To align with the PL emission band of the monolayer WSe$_2$, the membrane thickness was designed and fabricated to be 180 nm through a controlled etching process prior to the transfer of WSe$_2$. **Figure 3** demonstrates the large PL enhancement from monolayer WSe$_2$ achieved by coupling to the membrane photonic platform. Figure 3a presents the PL emission spectra of WSe$_2$ under two conditions, with the red curve corresponding to WSe$_2$ placed on the membrane photonic platform, and the black curve (multiplied by 10 for clarity) representing WSe$_2$ on an unstructured SiN membrane. A significant enhancement in PL intensity is observed when the monolayer WSe$_2$ is coupled to the membrane photonic platform, which supports quasi-BICs that enable efficient light–matter interaction across a large area. The red curve is decomposed into Lorentzian components corresponding to individual quasi-BIC resonances at wavelengths 753.9, 762.0, and 776.7 nm, corresponding to BIC$_1$, BIC$_2$, and BIC$_3$, respectively. These experimentally observed resonance wavelengths are in good agreement with the simulated values of 752.3, 762.0, and 776.5 nm. The Q-factors of the fitted PL peaks for BIC$_1$, BIC$_2$, and BIC$_3$ are 54, 118, and 92, respectively. These values were obtained by performing Lorentzian curve fitting using an iterative approach, in which the fitting parameters were optimized until the chi-square value converged. The inset compares the PL of WSe$_2$ on a 180-nm thick unstructured SiN membrane with that on a 180-nm thick SiN film supported by a silicon (Si) substrate, where the significantly enhanced PL observed in the unstructured membrane sample highlights the critical impact of substrate-induced field leakage on emission efficiency (see Figure S3). This confirms the advantage of using a freestanding membrane to suppress substrate losses and enhance optical confinement. Figure 3b shows the PL enhancement factor, defined as the ratio of the emission intensity from the membrane photonic platform to that from an unstructured SiN membrane. A peak enhancement factor of 1158 is observed near the quasi-BIC resonances, indicating that the high-Q modes of the membrane photonic platform significantly amplify the WSe$_2$ emission. Figures 3c and 3e display spatial emission intensity maps of WSe$_2$ PL across a 128 × 128-pixel grid for BIC$_2$ and BIC$_3$, respectively. The emission is uniformly distributed over the entire 450 × 450 μm² membrane photonic platform, confirming that the enhancement effect is robust and not confined to localized hotspots. Note that a weak-emission spot at the center of the platform is attributed to laser-induced damage during testing. Figures 3d and 3f present the spatial maps of the maximum PL wavelength for BIC$_2$ and BIC$_3$, respectively. These maps reveal well-defined, mode-specific spectral features that are consistent across the membrane photonic platform, indicating uniform resonance behavior and effective spectral tunability. Statistical analysis yields mean peak wavelengths of 758.3 nm for BIC$_2$ and 773.9 nm for BIC$_3$, with standard deviations of only 0.73 nm and 0.77 nm, respectively, indicating excellent spectral uniformity across the membrane. The small resonance wavelength variation further confirms the stability and consistency of the quasi-BICs across the entire platform.

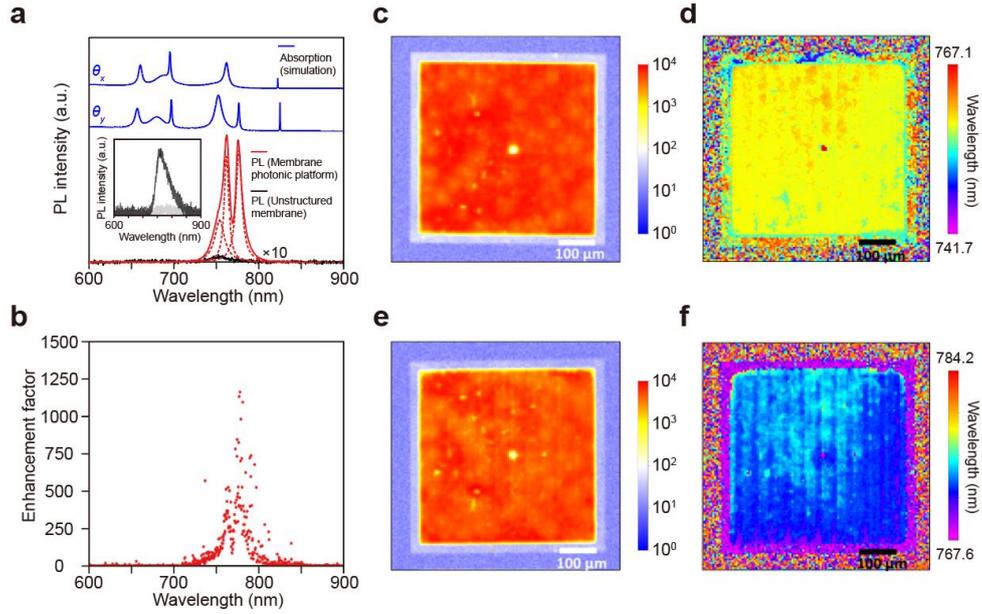

**Figure 3.** Large-area PL enhancement enabled by integration of the membrane photonic platform with monolayer WSe$_2$. a) PL emission spectra of monolayer WSe$_2$ on the photonic platform (red curve) and an unstructured SiN membrane (black curve, multiplied by 10 for clarity). The spectra are deconvoluted into individual Lorentzian peaks (dashed curves) with the overall fitting indicating BIC$_1$, BIC$_2$, and BIC$_3$ shown as a red dashed curve. Simulated absorption spectral (blue curve) of the 0.73 nm monolayer of WSe$_2$ on a membrane photonic platform, illuminated with *x*-polarized light at an incident angle of 4°, along the *x*-direction and *y*-direction. The inset shows a comparison of the PL spectra of WSe$_2$ on the unstructured membrane (black curve) and a SiN thin film supported by a Si substrate (gray curve). b) The enhancement factor determined by calculating the ratio of the PL intensity of WSe$_2$ on the photonic platform to that on the unstructured SiN membrane. c) and e) Spatial emission intensity maps for the BIC$_2$ and BIC$_3$, respectively, shown on a logarithmic scale. d) and f) Spatially resolved maps of the PL peak wavelengths for BIC$_2$ and BIC$_3$, respectively. The scale bars are 100 μm. The spatial maps consist of a 128×128-pixel grid.

Table 1 summarizes representative studies reporting PL enhancement in monolayer TMDs using large-area photonic platforms for enhanced light–matter interactions.[6-19] While photonic-crystal cavities or single plasmonic/photonic cavity-based localized platforms typically offer strong field confinement, they are limited in spatial coverage and scalability (see Table S1). In contrast, photonic platforms based on periodic structures engineered for large-area monolayer integration can sustain high-quality resonances over extended regions, making them highly compatible with wafer-scale CVD-grown TMDs. These features enable spatially uniform PL enhancement across large-area 2D materials. As shown in Table 1, most previous reports demonstrate enhancement factors on the order of $10^1$ to $10^2$, and values exceeding 1000 have seldom been reported. In comparison, our membrane photonic platform achieves a PL enhancement factor of approximately 1158, which is among the highest reported values for photonic platforms based on periodic structures. In addition to its high magnitude, the enhancement is spatially homogeneous and continuous over a 450 μm × 450 μm area, representing the large area for PL enhancement in monolayer TMDs. This performance is enabled by a photonic platform based on a suspended membrane architecture that supports quasi-BIC resonances, ensuring robust and uniform light–matter interaction across extended regions of monolayer TMDs. It is noted that the measured Q-factor of 118 and 92 enables precise spectral control, making the platform particularly advantageous for large-scale 2D material photonics.

**Table 1.** Photoluminescence enhancement in 2D TMDs enabled by photonic platforms at room temperature.

| TMD/Method | Structure | Resonance | $EF_{PL}$ | Effective area | Q | Ref. Year |
|---|---|---|---|---|---|---|
| $MoS_2$/CVD | Ag disk array | Plasmonic | 12 | 25 × 25 μm² | 14 | [6] 2015 |
| $WSe_2$/Exfoliation | Si grating waveguide | Photonic waveguide | 8 | 5 × 5 μm² | 35 | [7] 2017 |
| $WSe_2$/CVD | $Si_3N_4$ Bragg grating | Photonic waveguide | 2 | 2 × 2 μm² | 50 | [8] 2018 |
| $WS_2$/Exfoliation | Si bar pairs | BIC | 5 | 40 × 16 μm² | 56 | [9] 2020 |
| $WS_2$/CVD | $Si_3N_4$ grating | Photonic waveguide | 300 | - | 31 | [10] 2020 |
| $WS_2$/CVD | Ag disc array | Plasmonic | 16 | 13 × 13 μm² | - | [11] 2021 |
| $MoS_2$/CVD | Au sphere array | Plasmonic | 12 | $10^4 × 10^4$ μm² | 15 | [12] 2021 |
| $WSe_2$/Exfoliation | $TiO_2$ cuboid array | Mie resonance | 170 | 17 × 8 μm² | | [13] 2021 |
| $MoS_2$/CVD | Au rod array | Plasmonic | 60 | 30 × 60 μm² | 15 | [14] 2022 |
| $WS_2$/CVD | $Si_3N_4$ hole array | BIC | 10 | 500 × 500 μm² | - | [15] 2024 |
| $WS_2$/CVD | Au-Si hole array on Au film | Plasmonic–Photonic | 1030 | 960 × 600 μm² | 36 | [16] 2024 |
| $WSe_2$/CVD | Si pellet array | Photonic waveguide | 300 | 1200 × 600 μm² | 26 | [17] 2024 |
| WSe$_2$/ Exfoliation | SiN hole-array membrane | BIC | 40 | 20 × 20 µm$^2$ | - | [18] 2024 |
| WS$_2$/ Exfoliation | Si$_3$N$_4$ hole array | BIC | 5 | 100 × 80 µm$^2$ | 5 | [19] 2025 |
| **WSe$_2$/ CVD** | **SiN hole-array membrane** | **BIC** | **1158** | **450 × 450 µm$^2$** | **92 / 118** | **This work** |

## 5. Large-Area Nonlinear Optical Enhancement in Monolayer WSe$_2$ via Quasi-BIC Modes

In addition to enhancing PL, the platform exploits BIC resonances to amplify nonlinear optical processes. **Figure 4** illustrates the enhanced SHG from the monolayer WSe$_2$ coupled to the membrane photonic platform. To ensure that the BIC resonance is positioned within the 800 nm laser pump band, the membrane thickness was designed and fabricated to be 200 nm prior to the transfer of WSe$_2$, as predicted by the simulation results in Figures 2b and 2c. As shown in Figure 4a, the SHG signal from WSe$_2$ on the membrane photonic platform is significantly stronger than that from WSe$_2$ on an unstructured SiN membrane. The inset shows the PL spectrum of WSe$_2$ coupled to the membrane platform, featuring two peaks, the superimposed peak from BIC$_1$ and BIC$_2$ (787.1 nm) and the peak of BIC$_3$ (806.8 nm). Compared to the substrate-supported case, the unstructured SiN membrane already yields a stronger SHG response, as shown in Figure S3, highlighting the suppression of substrate-induced field leakage in the freestanding membrane configuration. The substantially higher SHG enhancement observed in the membrane photonic platform is attributed to the strong local field enhancement within the monolayer WSe$_2$, enabled by the quasi-BIC resonances supported by the membrane photonic platform. The peak enhancement factor was calculated to be 378. It is noted that, due to the broadband nature of the fs pulsed pump laser, the SHG spectrum reveals two narrowband resonances, providing direct evidence that the quasi-BIC resonances contribute to the nonlinear optical processes. Multiple well-resolved SHG peaks with narrow linewidths, originating from high-Q quasi-BICs, have not been reported in earlier studies.[9, 20-22, 30-34]

Spatial mapping of the SHG emission wavelength corresponding to the maximum intensity (Figure 4c) demonstrates a uniform distribution of the superimposed peak of BIC$_1$ and BIC$_2$ across the entire membrane photonic platform, highlighting the robustness and scalability of the proposed device. Statistical analysis shows a mean peak wavelength of 397.9 nm with a standard deviation of only 1.31 nm, indicating high spatial uniformity. A distinct red square corresponding to the unstructured SiN membrane region is observed, where the SHG signal shifts to longer wavelengths. Figures 4d–f present the power-dependent SHG measurements for monolayer WSe$_2$ placed on the Si substrate, the unstructured SiN membrane, and the membrane photonic platform, respectively. The observed power ($P$) dependence of the SHG intensity ($I_{SHG}$) follows the expected quadratic scaling ($I_{SHG} \propto P^2$), yielding a slope of ~2 in the log–log plot, as shown in the insets. This confirms the preservation of the intrinsic second-order nonlinear optical response of monolayer WSe$_2$. The enhanced local fields enabled by quasi-BIC resonances further promote efficient SHG generation, demonstrating the platform's ability to significantly amplify nonlinear optical processes over large areas while retaining the fundamental second-order nature of the material response.

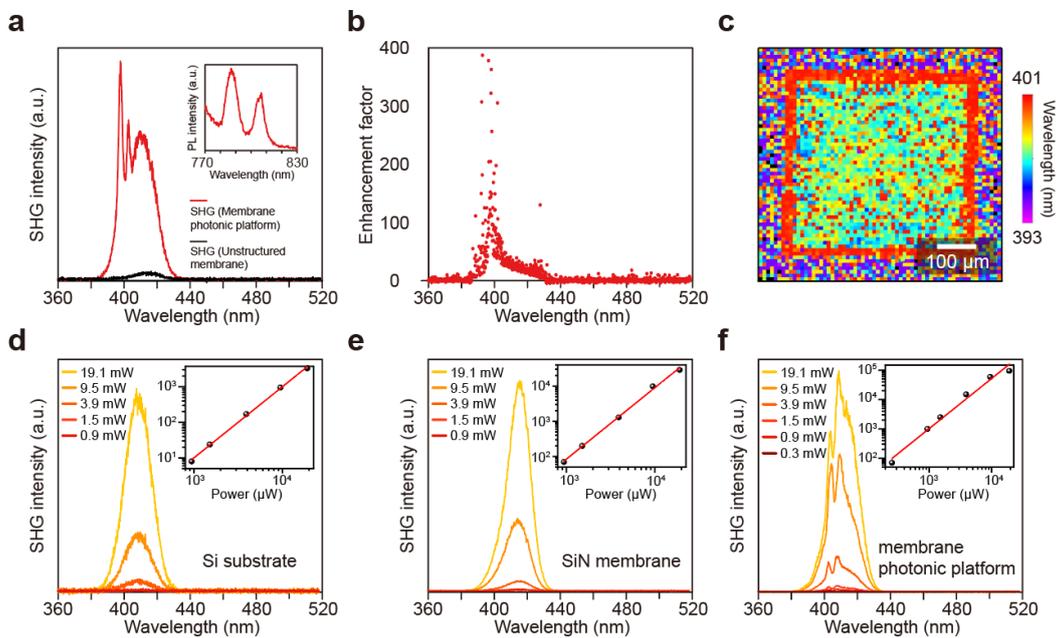

**Figure 4.** Large-area SHG enhancement from monolayer WSe$_2$ enabled by the membrane photonic platform. a) SHG spectra of monolayer WSe$_2$ on the photonic platform (red curve) and on the unstructured SiN membrane (black curve). The inset shows the PL spectrum of WSe$_2$ on the membrane photonic platform at the same sample. b) The SHG enhancement factor was calculated as the ratio of the SHG intensity of monolayer WSe$_2$ on the membrane photonic platform to that on the unstructured membrane. c) Spatial SHG emission wavelength mapping of WSe$_2$ on the membrane photonic platform. d, e, f) SHG emission spectra of WSe$_2$ at varying excitation powers for the d) SiN thin film on a Si substrate, e) unstructured SiN membrane, and f) membrane photonic platform. The insets show the log-log plots of the integrated SHG intensity versus excitation power.

To further investigate the nonlinear optical characteristics, polarization-resolved SHG measurements were performed on the monolayer WSe$_2$ coupled to the membrane photonic platform. Figure S4 illustrates the corresponding measurement setup. An 800 nm (ω) pulsed laser is used as the excitation source, which is the same as in the previous section. After passing through a polarizer, the beam is linearly polarized and directed onto the sample. The generated second-harmonic (2ω) signal then passes through a polarization analyzer fixed at 90°, orthogonal to the incident polarization, thereby forming a cross-polarized detection configuration. By rotating the sample and recording the SHG intensity as a function of the rotation angle, the polarization-resolved SHG response is obtained, revealing the crystalline symmetry and nonlinear optical characteristics of the monolayer WSe$_2$. Figure 5 illustrates the mapping polarization-resolved SHG characterization of WSe$_2$ coupled to the membrane photonic platform. **Figure 5**a shows four SHG intensity mapping images acquired at rotation angles of 0°, 30°, 60°, and 90°. Pronounced spatial variations in SHG intensity are observed across the field of view, with clearly distinguishable bright and dark regions. These variations reflect differences in the local crystal orientation of individual WSe$_2$ domains, as the second-order nonlinear polarization depends on the projection of the incident electric field onto the optical axes. As the sample rotates, the intensity in each region changes periodically, revealing the angular dependence of the SHG response. The corresponding polar plot in Figure 5b shows a clear six-fold symmetry, consistent with the hexagonal lattice symmetry of monolayer WSe$_2$. Figure 5c displays the SHG intensity mapping at 0° rotation, display with a smaller intensity scale range, reveals the SHG signal from monolayer WSe$_2$ coupling to the membrane photonic platform, as well as from the surrounding Si substrate regions. The enhanced brightness in this image clearly reveals the spatial distribution of WSe$_2$ across both the membrane photonic platform and surrounding Si substrate areas.

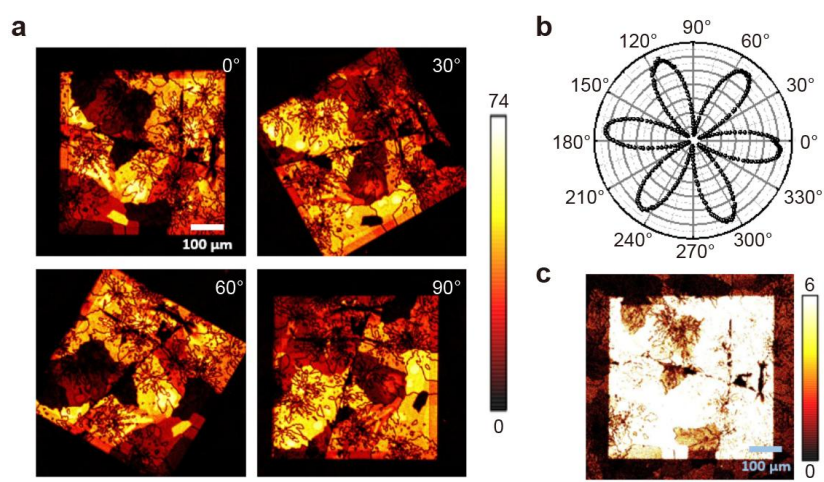

**Figure 5.** Polarization-resolved SHG enhancement mapping from monolayer $WSe_2$ enabled by the large-area membrane photonic platform. a) SHG intensity mapping of the monolayer $WSe_2$ coupling to the membrane photonic platform under sample rotation angles of 0°, 30°, 60°, and 90°. b) SHG intensity as a function of the sample rotation angle, measured for the monolayer $WSe_2$-coupled membrane photonic platform. c) SHG intensity mapping at 0° rotation with a smaller intensity scale range. Scale bars represent 100 μm.

Table 2 summarizes recent studies on SHG enhancement in monolayer TMDs using large-area photonic platforms.[9, 20-22] Consistent with the PL enhancement design approaches discussed earlier, our method is based on photonic platforms using periodic structures engineered for large-area monolayer integration. These periodic architectures are well-suited to sustain high-quality optical resonances across extended regions, making them highly compatible with wafer-scale CVD-grown TMDs. Our membrane photonic platform exhibits an SHG enhancement factor of 378 over a continuous and spatially homogeneous 450 μm × 450 μm area, combining high SHG enhancement and spatial uniformity, and demonstrating a clear and notable advantage over previously reported results. The large-area enhancement supports efficient nonlinear conversion and enables SHG-based mapping of crystal orientation and grain boundaries in monolayer $WSe_2$. This combination of strong optical enhancement and spatially extensive imaging capability offers a valuable pathway for scalable characterization and integration of 2D semiconductors.

**Table 2.** Second-harmonic generation enhancement in 2D TMDs enabled by photonic platforms at room temperature

| TMD/ Method | Structure | Resonance | Pump wavelength | $EF_{SHG}$ | Effective area | Ref. Year |
|---|---|---|---|---|---|---|
| $WS_2$/ Exfoliation | Si bar pairs | BIC | 832 | 1140 | 40 × 16 μm² | [9] 2020 |
| $MoS2$/ CVD | Si bar pairs | BIC | 850 | 35 | 15 × 10 μm² | [20] 2020 |
| $MoS2$/ CVD | SiN on Ag hole-array membrane | Plasmonic | 870 | 1527 | 22 × 22 μm² | [21] 2021 |
| $MoS2$/ CVD | $SiO_2$/$TiO_2$ DBR pair | FP cavity | 807 | 16 | 5 × 2 μm² | [22] 2023 |
| $WSe_2$/ CVD | **SiN hole-array membrane** | **BIC** | **800** | **378** | **450 × 450 um²** | **This work** |

## 6. Conclusion

In summary, we have developed a large-area freestanding membrane photonic platform that achieves significant enhancement and uniformity in light–matter interactions with monolayer WSe$_2$. By utilizing high-Q quasi-BIC resonances in a substrate-free photonic membrane, the platform delivers PL and SHG enhancement factors of 1158 and 378, respectively, uniformly across a 450 × 450 μm$^2$ area. Notably, the SHG spectrum reveals multiple clearly resolved and narrow-linewidth peaks originating from distinct quasi-BIC resonances, providing direct spectral evidence of resonantly enhanced nonlinear coupling. In addition, the large-area SHG enhancement enables polarization-resolved SHG mapping of crystal orientation and grain boundaries in monolayer WSe$_2$, establishing a promising strategy for spatially extensive structural characterization of 2D materials. This architecture uniquely integrates strong electromagnetic field confinement, spectral tunability, and large-area compatibility, effectively overcoming limitations associated with localized nanocavity approaches. The demonstrated performance establishes a robust and scalable platform for integrating 2D materials into advanced optoelectronic and nanophotonic systems, and opens new avenues for future applications in nonlinear optics and quantum photonics.

## 7. Experimental Section/Methods

*Numerical Simulations*

The far-field absorption spectra of the WSe$_2$-coupled membrane photonic platform, as well as the near-field electric field distributions of the resonance modes, were computed using the rigorous coupled-wave analysis (DiffractMOD, RSoft Design Group, USA). All simulations were performed under periodic boundary conditions in the *x*- and *y*-axes and perfectly matched-layer conditions in the *z*-axis, with plane-wave light incidence and the incident angle along the *z*-axis. The electric field **E** is normalized by the electric field amplitude of the incident light. The electric energy density is defined as $U_E = ½∫\text{Re}[\varepsilon(\mathbf{r}')]|\mathbf{E}|^2 \, dV$, where **E** is the electric field, ε is the spatially dependent permittivity, and V is the volume of the simulation grid.

*Preparation of Monolayer TMD*

The AACVD synthesis of monolayer WSe$_2$ film involved a horizontal tube furnace setup. In this process, a quartz tube was used to house the solid precursors and substrate. H$_2$WO$_4$ and Se powder were used as the tungsten and selenium sources, respectively, while KOH powder served as the growth promoter. The clean c-plane sapphire substrate was placed face-down near the H$_2$WO$_4$ source to facilitate nucleation and lateral growth. An argon carrier gas was introduced from the gas inlet, transporting the precursors toward the substrate. The furnace temperature was ramped up from room temperature to 975˚C. After a 10-minute reaction period, the furnace was allowed to cool naturally to room temperature. During the growth, the temperature profile, carrier gas flow rate, and precursor placement were carefully controlled to optimize nucleation density, domain size, and crystal quality.

*Transfer of monolayer TMD onto the membrane photonic platform*

We employed an EVA-assisted wet transfer method to achieve high-quality material transfer. First, a 4% EVA solution was heated to 80˚C and spin-coated onto the $WSe_2$ sample at 8000 rpm for 60 seconds to form a uniform supporting layer. After baking at 80˚C for 10 minutes to solidify the EVA, the sample was immersed in 3M NaOH to etch the sapphire substrate, allowing the $WSe_2$ film to detach and float on the alkaline solution. The $WSe_2$ was then carefully scooped up and rinsed three times with deionized water to remove any residual alkali. Before transfer, the target membrane photonic platform was treated with oxygen plasma to clean the surface and enhance hydrophilicity ($O_2$ flow: 70 sccm, pressure: $7 \times 10^{-1}$ torr, 30 W, 5 minutes). The treated membrane platform was then used to pick up the $WSe_2$ film, which was baked on a hot plate at 80˚C for 10 minutes to ensure adhesion. Finally, the sample was immersed in xylene for 10 minutes to dissolve the EVA support layer, completing the transfer process. The SiN membranes are provided by Ted-Pella Inc. and further nanofabricated and utilized in this work. To precisely tune the BIC resonance wavelengths to match the PL emission peak of $WSe_2$, the membrane thickness was gradually reduced using a low-damage reactive ion etching (RIE) process, enabling highly accurate thickness control.

*Optical Characterization*

PL measurements were conducted using a continuous-wave He-Ne laser at 633 nm as the excitation source, filtered by a 633 nm laser line filter. The micro-PL signal was measured using an Olympus 4× PlanN objective lens (NA = 0.10). The emission was filtered using a 633 nm long-pass filter, before being coupled into a mid-resolution spectrometer equipped with a ruled reflective diffraction grating (300 lines/mm, 762 nm blaze) through a Thorlabs 105 μm core diameter optical fiber.

SHG measurements were performed using a Ti:Sapphire femtosecond laser source at 800 nm. The laser beam was focused onto the sample and the generated SHG signal was collected using an Olympus 4× PlanN objective lens (NA = 0.10). The SHG spectra were acquired with a high resolution spectrometer equipped with a ruled reflective diffraction grating (300 lines/mm, 500 nm blaze). The SHG intensity imaging was performed using a custom-built laser scanning confocal microscopy system with an Olympus 4× PlanN objective lens (NA = 0.10). Polarization-resolved SHG measurements were performed using an 800 nm ($\omega$) femtosecond laser as the excitation source. The linearly polarized light was incident on the TMDs sample. Through nonlinear optical interactions, the sample generates a second-harmonic signal at 400 nm ($2\omega$). Notably, the SHG emission was initially unpolarized. When the SHG emission passed through an analyzer which is orthogonal to the incident laser polarization, the detection configuration was cross-polarized. By rotating the sample and measuring the corresponding SHG intensity, the polarization-resolved SHG response of the material could be obtained. Schematic illustrations are shown in Figure S4.

**Acknowledgements**

This work was supported by JSPS KAKENHI Grant Number JP23K26155, JP25KF0083. A part of this work was supported by Advanced Research Infrastructure for Materials and Nanotechnology in Japan (ARIM) of the Ministry of Education, Culture, Sports, Science and Technology (MEXT) (Proposal Number JPMXP1225NM5090). Financial support by the Center of Atomic Initiative for New Materials (AI-Mat), National Taiwan University, from the Featured Areas Research Center Program within the framework of the Higher Education Sprout Project by the Ministry of Education in Taiwan (Grant Number 111L900801), is also acknowledged. F. L. Hsieh and C. Z. Deng contributed equally to this work.

**Data Availability Statement**

The data that support the findings of this study are available from the corresponding author upon reasonable request.

A freestanding membrane photonic platform is demonstrated to achieve strong photoluminescence and second-harmonic generation enhancement in monolayer $WSe_2$ via quasi-bound states in the continuum. The large-area and uniform enhancement enables multiresonant nonlinear emission and polarization-resolved structural mapping, offering a scalable route toward integrated nanophotonics, optoelectronics, quantum photonic devices, and large-scale structural characterization of two-dimensional semiconductors.

**Large-Area Photonic Membranes Achieving Uniform and Strong Enhancement of Photoluminescence and Second-Harmonic Generation in Monolayer $WSe_2$**

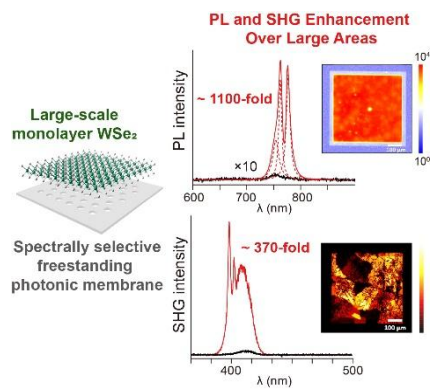

Supporting Information

**Large-Area Photonic Membranes Achieving Uniform and Strong Enhancement of Photoluminescence and Second-Harmonic Generation in Monolayer WSe$_2$**

*‡Fong-Liang Hsieh, ‡Chih-Zong Deng, Shao-Ku Huang, Tsung-Hsin Liu, Chun-Hao Chiang, Che-Lun Lee, Man-Hong Lai, Jui-Han Fu, Vincent Tung, Yu-Ming Chang, Chun-Wei Chen\*, and Ya-Lun Ho\**

Figure S1 presents the simulated absorption spectra of a 0.6 nm monolayer $WSe_2$ on the membrane photonic platform as a function of incident angle θ, varied from 0° to 5°. This narrow angular range is chosen to isolate symmetry-protected bound states in the continuum (BICs), which are expected to occur at the Γ point. At normal incidence ($\theta = 0°$), the spectrum exhibits optical modes with theoretically infinite Q-factors—a defining characteristic of ideal BICs—enabling clear differentiation from guided modes.

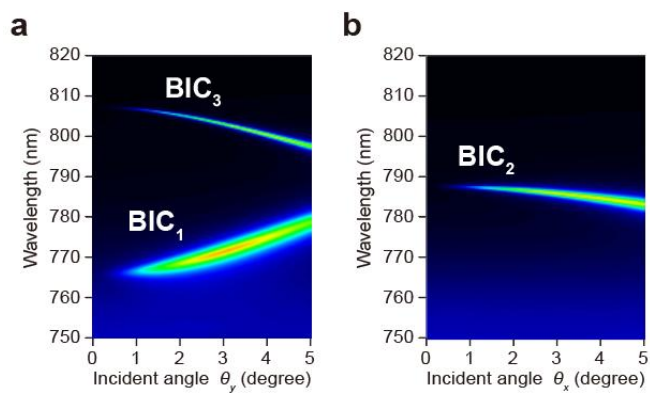

**Figure S1.** Simulated absorption spectra as a function of incident angle for a 0.6 nm monolayer of $WSe_2$ on a membrane photonic platform, illuminated with *x*-polarized light at an incident along the a) *y*-direction and b) *x*-direction.

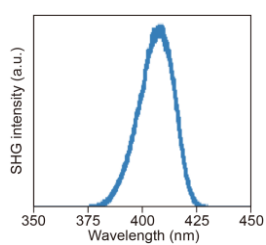

**Figure S2.** SHG spectra of as-grown $WSe_2$ are shown in the figure. SHG spectrum displays a peak at 400 nm, which is half of the excitation wavelength and arises from the efficient second-order nonlinear optical response of the non-centrosymmetric $WSe_2$ monolayer.

Figure S3 shows the enhancement factors for a) PL and b) SHG, defined as the ratio of emission intensity from the unstructured SiN membrane to that from a SiN film supported by a silicon (Si) substrate, using the PL and SHG signals of monolayer $WSe_2$. The significantly reduced emission observed in the substrate-supported configuration highlights the critical impact of substrate-induced field leakage on optical efficiency. In contrast, the freestanding membrane configuration effectively suppresses this leakage, resulting in enhanced light–matter interaction.

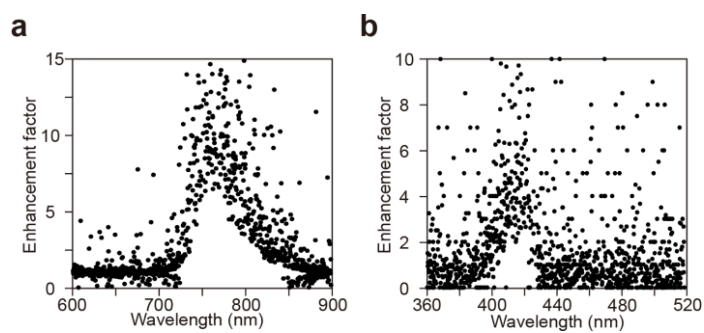

**Figure S3.** a) PL and b) SHG enhancement factors, defined as the ratio of the emission intensity of monolayer $WSe_2$ on the plain membrane to that on a SiN film supported by a Si substrate.

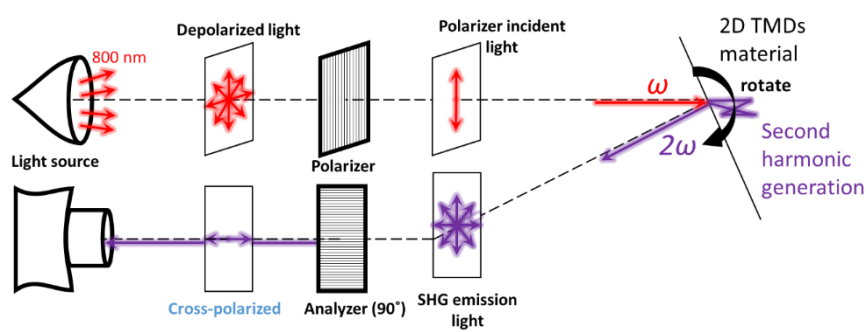

**Figure S4.** Schematic of the polarization-resolved SHG measurement setup. A pulsed laser at 800 nm (ω) passes through a polarizer to produce linearly polarized light, which is then focused onto the sample. The sample is mounted on a rotational stage to vary the incident polarization angle. The generated SHG signal at 400 nm (2ω) is collected in the cross-polarized configuration using an analyzer set at 90°, allowing angular-resolved SHG intensity measurements.

**Table S1.** Photoluminescence enhancement in 2D TMDs enabled by independent cavities at room temperature

| TMD/ Method | Structure | Resonance | $EF_{PL}$ | $EF_{Norm.\,PL}$[1] | Effective area | Q | Ref./ Year |
|---|---|---|---|---|---|---|---|
| $MoS_2$/ CVD | Au Rod | Plasmonic | 6.5 | - | 25 × 57 $nm^2$ | 23 | [1] 2015 |
| $WSe_2$/ CVD | Au Trench | Plasmonic | 40 | 20000 | 12 × 12 $nm^2$ | 24 | [2] 2015 |
| $MS_2$/ CVD | Ag Wire | Plasmonic | 38 | - | 120 × 5000 $nm^2$ | - | [3] 2017 |
| $WSe_2$/ Exfoliation | Au cube | Plasmonic | 13 | - | 110 × 110 $nm^2$ | - | [4] 2018 |
| $MoS_2$/ CVD | Ag cube | Plasmonic | 211 | 6100 | 75 × 75 $nm^2$ | 12 | [5] 2018 |
| $WSe_2$/ Exfoliation | GaP dimer | Mie resonance | 50 | $10^4$ | 400 × 200 $nm^2$ | 33 | [6] 2019 |
| $MoS_2$/ CVD | Au particle | Plasmonic | 7.7 | - | π × 200 × 200 $nm^2$ | 19 | [7] 2020 |
| $MoS_2$/ Exfoliation | Au dimer | Plasmonic | 32 | 1350 | 280 × 140 $nm^2$ | 13 | [8] 2021 |
| $MoS_2$/ CVD | Ag wire dimer | Plasmonic | 14.5 | - | 100 × 1000 $nm^2$ | 18 | [9] 2022 |
| $WS_2$/ Exfoliation | Au disk | Plasmonic | 4 | - | π × 1100 × 1100 $nm^2$ | 47 | [10] 2023 |
| $WSe_2$/ Exfoliation | Au cube | Plasmonic | 7 | 1800 | 99 × 99 $nm^2$ | 42 | [11] 2024 |

[1]$EF_{Norm.PL}$ is defined as $(I_{cav} - I_0)S_0/I_0 S_{cav}$, where $I_{cav}$ and $I_0$ are the PL intensities of the TMD when coupled and uncoupled to the cavity, respectively. $S_{cav}$ and $S_0$ are the areas of the cavity and the laser spot, respectively.

**Table S2.** Second-harmonic generation enhancement in 2D TMDs enabled by independent cavities at room temperature

| TMD/ Method | Structure | Resonance | $\lambda_{pump}$ (nm) | $EF_{SHG}$ | $EF_{Norm.\ SHG}$[1] | Effective area | Ref./ Year |
|---|---|---|---|---|---|---|---|
| WSe$_2$/ Exfoliation | Si hole structure | Photonic-crystal cavity | 1515 | 200 | - | 900 × 3900 nm$^2$ | [1] 2020 |
| WS$_2$/ CVD | Ag cube | Plasmonic | 820 | 3 | 300 | 75 × 75 nm$^2$ | [2] 2021 |
| MoS$_2$-WS$_2$/ Exfoliation | Hole in SiO$_2$/Si | FP cavity | 800 | 1713 | - | 3000 × 1600 nm$^2$ | [3] 2022 |
| MoSe$_2$/ Exfoliation | Au nanocircuit | Plasmonic | 1560 | 13.8 | - | 4000 × 130 nm$^2$ | [4] 2023 |
| WSe$_2$/ Exfoliation | Si$_3$N$_4$ nanocavity | Photonic waveguide | 804 | - | 8.95 × 10$^3$ | 20000 × 220 nm$^2$ | [5] 2024 |

[1]$EF_{Norm.SHG}$ is defined as $(I_{cav} - I_0)S_0/I_0 S_{cav}$, where $I_{cav}$ and $I_0$ are the SHG intensities of the TMD when coupled and uncoupled to the cavity, respectively. $S_{cav}$ and $S_0$ are the areas of the cavity and the laser spot, respectively.